\documentclass{JMLFS}
\usepackage{natbib}
\usepackage{url}
\usepackage{hyperref}
\journal{JAIS-ID}
\vol{2023}
\usepackage{lineno}

\received{1 August 2023}

\def\be{\begin{equation}}
\def\ee{\end{equation}}
\def\bea{\begin{eqnarray}}
\def\eea{\end{eqnarray}}

\begin{document}

\title{Design Challenges for a Future Liquid Xenon Observatory}

\author{Abigail Kopec\auno{1}}
\address{$^1$ Department of Physics \& Astronomy, Bucknell University, Lewisburg, PA, USA 17837}

\address{Email: amk029@bucknell.edu}

\begin{abstract}
An ultimate liquid xenon experiment would be limited in its dark matter science reach by irreducible neutrino backgrounds, which are an exciting signal in their own right. To achieve such sensitivity, other backgrounds that currently plague these detectors must be better mitigated, and extreme care must be taken in the design and construction phases. A 100-tonne xenon target is compelling to search for weakly interacting massive particle dark matter, and has capabilities to study coherent elastic neutrino-nucleus scattering and search for neutrinoless double-beta decay signatures. Historically, liquid xenon time projection chambers have scaled to larger target masses with great success. This paper gives an overview of challenges that need to be met for the next generation of detector to obtain a kilotonne$\times$year exposure. Such tasks include the procurement and purification of xenon, radiopure and reliable detector components, sensitive outer detector vetoes, powerful data handling and analyses, and an ability to operate stably for timescales of over a decade. 

\end{abstract}

\maketitle


\section{Introduction}\label{intro}

Liquid xenon time projection chambers (LXe TPCs) have dominated searches for spin-independent dark matter in the form of weakly interacting massive particles (WIMPs). Current experiments, including LUX-ZEPLIN (LZ)~\cite{LZ:2022ufs}, XENONnT~\cite{XENON:2023sxq} and PANDAX-4T~\cite{PandaX:2021bab}, are expected to continue this legacy of excellence, employing world-leading detection capabilities to search for rare particle interactions over the next several years~\cite{LZ:2018qzl,XENON:2020kmp}. To optimally exploit the dark matter science potential of this technology, a liquid xenon observatory is being proposed~\cite{Aalbers:2022dzr}. The DARWIN Collaboration has made significant progress toward the next upgraded detector~\cite{DARWIN:2016hyl,DARWIN:2022ewl} and the new XENON-LUX-ZEPLIN-DARWIN (XLZD) Consortium aims to achieve it with community support~\cite{Cooley:2022ufh}. Optimally, coherent elastic neutrino-nucleus scattering (CEvNS) backgrounds from irreducible solar and atmospheric neutrinos would be the only limitations to such a detector's dark matter sensitivity. A target exposure of a kilotonne$\times$year also has the discovery potential for rare neutrino physics, such as the neutrinoless double-beta decay process~\cite{DARWIN:2020jme}, and observing the natural neutrino backgrounds and their properties~\cite{Akerib:2022ort}. The broad science reach enables the future detector to be an observatory rather than just the next liquid xenon dark matter experiment~\cite{Aalbers:2022dzr}. Realistically, discriminating between rare particle interactions and backgrounds--such as radioactive decays or mistaken events from random signal coincidences--makes these experimental goals more challenging. 

There are several merits of LXe TPCs. A particle interacting in the target volume efficiently transfers energy to create excimers, some of which decay within nanoseconds by emitting vacuum ultraviolet scintillation photons~\cite{Hogenbirk:2018zwf}. Other excited xenon atoms ionize and the electrons can be drifted in an electric field to an amplification region where they produce a secondary signal. A small portion of the energy is lost as heat, which is not directly measurable. In the common dual-phase LXe TPCs, a drift electric field is applied to the bulk liquid xenon to draw ionization electrons up to the gaseous xenon. A stronger field is applied across the liquid surface to extract electrons and accelerate them to the point of creating proportional scintillation in the gas. Therefore, particle interactions are detected via two correlated signals: the prompt scintillation (S1) and the delayed ionization electrons' proportional scintillation (S2). 

Three electrodes produce the two electric fields: the cathode at the bottom, the gate just below the liquid surface and the anode in the gas just above the surface. A field cage of stacked field-shaping rings around the liquid xenon target volume are set at incremental voltages between the cathode and gate to reduce distortions in the field at the edge of the detector. At a constant drift field, the electrons move through the liquid at a constant terminal velocity. The drift time after the S1 to the detection of the S2 provides a measurement of the depth $(z)$ at which the interaction occurred. Geometrically, the field cage leaves the top and bottom faces of the detector as the best locations for photosensors that detect the scintillation photons. This placement, the usage of photomultiplier tubes (PMTs) as the photosensors, and the transparency of the electrodes combine to affect the light collection efficiency. The ratio of S1 photons measured relative to photons produced (g1) is historically below 20\%. Of the current generation, XENONnT has achieved about 15\%~\cite{XENON:2023sxq}. 

The ionization electron detection (S2 detection) relies on the extraction field between the gate and the anode. The electroluminescence scintillation photons are highly localized in the photosensors directly above the extraction location, leading to excellent $(x,y)$ position reconstruction. With a dielectric constant of 1.85~\cite{Xu:2019dqb} in the liquid xenon and close to 1 in the gas, there is a potential difference at the liquid-gas interface, usually modeled as a Schottky Barrier and dependent on the extraction field~\cite{Gushchin:1979}. Although it is difficult to define the point at which all electrons have been extracted, the extraction efficiency starts to saturate above 7.4~kV/cm in the liquid~\cite{Xu:2019dqb}. The distance from the liquid surface to the anode, the gas pressure, and the electric field determine how much proportional scintillation is produced by the ionization electrons~\cite{Aprile:2008bga}. By additionally considering geometry and PMT characteristics, single extracted electrons produce signals of hundreds of photons and have a high detection efficiency. The LZ detector has achieved a single electron gain of 47 photoelectrons (PE) per electron with 80\% extraction efficiency~\cite{LZ:2022ufs}. At MeV interaction energy scales, there are roughly $10^5$~electrons contributing to the S2, which can lead to PMT and data acquisition saturation~\cite{XENON:2020iwh}.

Requiring a time-correlated S1 and S2 pair strongly suppresses backgrounds and provides information about the interaction. An interaction's deposited energy can be determined with an energy resolution of a few percent, depending on the corresponding quanta of photons and electrons measured~\cite{XENON:2020iwh}. Most backgrounds are due to radiation from the walls, which does not penetrate more than a few centimeters into the target. The position reconstruction enables fiducialization to select the cleanest xenon in the detector's center with the lowest rates of radioactive backgrounds. 

The type of interaction can also be determined. Electronic recoils (ERs) have a higher ratio of ionization electrons to scintillation photons (evident in an S2:S1 ratio) compared to nuclear recoils (NRs)~\cite{Hitachi:1983zz}. A recoiling nucleus transfers its energy to the surrounding xenon differently than a recoiling electron. WIMP dark matter is expected to produce NRs. Generally, a stronger drift field affects ERs more, suppressing recombination by stripping electrons away and thus increasing the S2:S1 ratio. For a small range of drift fields, the ER and NR populations are most distinguishable in S2 vs S1 parameter space. This optimal drift field is taken to be between 200-350~V/cm resulting in a leakage fraction of $7.3\times10^{-4}$ ER events in an NR range spanning 2-65~keV interactions~\cite{LUX:2020car,Szydagis:2022ikv}.

This work outlines several considerations for, and current limitations to, realizing a kilotonne$\times$year exposure with liquid xenon. It explores material and equipment considerations to build the next generation detector, while remarking on the limitations in the technology at the current generation. Significant progress has already been made in each successive experiment, and research and development is underway to ensure success.  

\section{Future Detector Considerations}\label{materials}

The atmospheric concentration of xenon is roughly 90 parts per billion by volume and, at the time of writing, about 60 tonnes are produced annually in a \$250 million USD industry~\cite{business:2022}. It is optimistically feasible for the XLZD consortium to collect about 100~tonnes of xenon over the coming years. 

In addition to the expense of obtaining xenon, the detector components must perform optimally without contaminating the xenon. Radioactive impurities that create background ER signals must be removed. The purity of the xenon regarding non-radioactive contaminants (such as oxygen) is critical at larger scales, since electronegative species capture drifting electrons and thus reduce the eventual S2 signals.  Accidentally coincident S1s and S2s are a major problem that necessitates a deeper understanding of isolated S1 and S2 rates, and creative analyses. Calibrations must be carefully considered to acquire necessary statistics and avoid uselessly high event rates. PMTs and electrodes are fragile and prone to failures.  With such a detector, a kilotonne$\times$year exposure would need to accumulate for over a decade, which requires the detector to be incredibly durable.

\subsection{Xenon}\label{sub_m:xenon}
Xenon currently costs a few million USD per tonne, and buying tens of tonnes of xenon has large implications for the market. Other significant uses for xenon include satellite propulsion, lighting and imaging, and semiconductor and electronics manufacturing~\cite{business:2022}. These applications all consume the xenon, and it eventually returns to the environment mixed with other byproducts or is lost in space. Liquid xenon particle detectors retain xenon over the lifetime of the experiments. Therefore, consumer companies could lend xenon to an experiment and have it returned in better condition with better purity after the experiment is decommissioned. 

Great care is needed when procuring xenon to optimize budgets and timelines. In addition to combining existing research experiments' xenon inventory, principal investigators of a future experiment should accumulate xenon over time. Buying in bulk may not be as cost-effective as spreading the responsibility over many institutions and countries. 100~tonnes of xenon is still worth buying from commercial producers rather than building a dedicated plant to produce it. There may be several viable strategies to accumulate xenon, but the timescale to obtain a kilotonne$\times$year exposure in liquid xenon requires advanced planning over several years, even before a decade of data-taking.

There are historic precedents for expensive community projects with decades-long timelines. Recently, detectors have not only grown in size, but also in the number of collaborators and overall lifetime. Over 500 individuals contributed to outlining the physics possibilities of a future xenon observatory~\cite{Aalbers:2022dzr}. Similar argon-based particle detectors with tens of tonnes of liquid argon are already being built. In particular, DarkSide-20k has outlined their procurement plan, starting with the extraction of underground argon in Colorado, USA, and processing it in Sardinia, Italy to remove the $^{39}$Ar isotope~\cite{DarkSide-20k:2021nia}. Although argon is plentiful and inexpensive, the facility required to remove the $^{39}$Ar requires a significant initial capital investment. In the context of contemporary experiments, this next generation liquid xenon detector is reasonable to fund, but care and time will need to be spent accumulating xenon in the small global market.   

\subsection{Radioactive Backgrounds}\label{sub_m:radioactive}

The natural isotopic abundances of xenon are particularly stable. Even the most common radioactive isotope, $^{136}$Xe, only undergoes a double-beta decay with a half-life of $2.165 \times 10^{21}$~years~\cite{EXO-200:2013xfn}. This double-beta decay process is of significant scientific interest for probing the nature of neutrinos. A future liquid xenon observatory is able to search for neutrinoless double-beta decay and evidence for Majorana neutrinos with a half-life on the order of $10^{27}$~years~\cite{DARWIN:2020jme}, which is competitive with dedicated neutrinoless double-beta decay experiments~\cite{Adams:2022jwx}. The half-life of the other main radioactive isotope, $^{124}$Xe, is so long that it was only recently measured to be $1.8 \times 10^{22}$~years in the XENON1T detector~\cite{XENON:2019dti}. In addition to natural isotopes, trace amounts of shorter-lived xenon isotopes have been observed, particularly after neutron NR calibrations due to neutron capture~\cite{XENON:2020rca}. These decay in a matter of days, and the next-generation xenon detector could run stably for years. If the detector behavior does not change over time, there is no need to take regular neutron calibrations that create these activated backgrounds. Xenon undergoes some activation by cosmic rays to produce $^{127}$Xe, but it has a half-life of 36.4~days and mostly disappears once the xenon is housed underground~\cite{Baudis:2015kqa}. With roughly 100~tonnes of xenon, detectable rates of cosmogenic activation will still happen depending on the underground site~\cite{DARWIN:2023uje}. Neutron and muon backgrounds will have to be constrained to both limit neutron NR backgrounds and characterize the production of radioactive nuclei in the xenon through muon spallation products and neutron capture. The topic of constraining neutrons and muons is further discussed in Section~\ref{sub_t:discrimination}.

The main radioactive backgrounds limiting xenon dark matter and neutrino physics sensitivity are from other noble elements mixed into the xenon, such as radon isotopes, $^{85}$Kr, and $^{37}$Ar. The current experiments have effectively eliminated $^{85}$Kr by processing the xenon, especially before it enters the system. XENONnT reuses the XENON1T krypton distillation column, which is able to remove $^{85}$Kr down to 0.36 ppt~\cite{XENON:2021fkt}. LZ has used an activated carbon trap that can reduce concentrations of noble element impurities down to similar levels~\cite{LUX:2016wel}. Cosmogenic activation of xenon stored above ground also produces $^{37}$Ar, which has a half-life of 35~days. The first results of LZ notably contain elevated levels of $^{127}$Xe and $^{37}$Ar~\cite{LZ:2022kml,LZ:2022ufs}, but they are expected to decay away over the lifetime of the experiment. The XENON program's krypton distillation column is also able to remove $^{37}$Ar with some system adjustments~\cite{XENON:2021fkt,XENON:2022ivg}. If the rates of cosmogenic activation even underground are noticeable in a future liquid xenon observatory, it is important to actively reduce $^{85}$Kr and $^{37}$Ar backgrounds.

\begin{figure}
    \centering
    \includegraphics[scale=0.5]{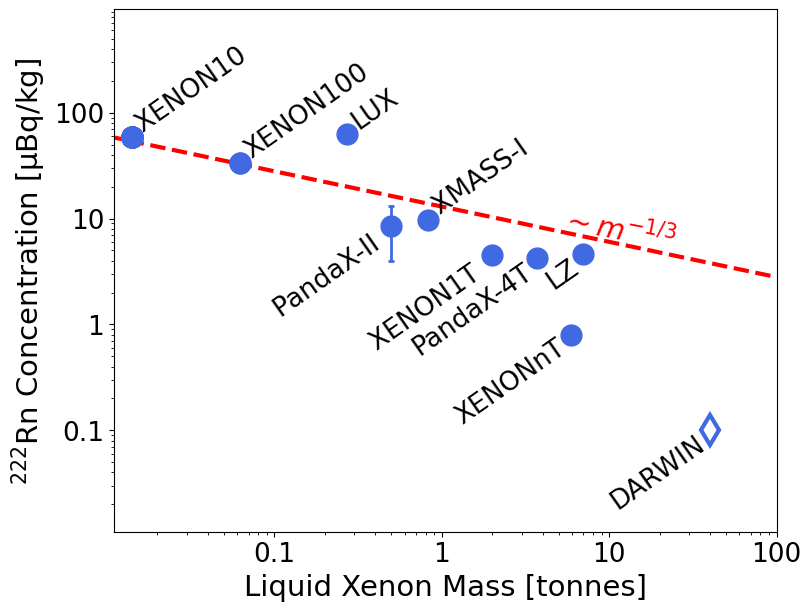}
    \caption{Achieved radon rates in LXe dark matter detectors (blue dots) and baseline proposed rate for an example 40~tonne volume of DARWIN (blue-outlined diamond)~\cite{DARWIN:2016hyl}. As the mass increases, a detector with an equal aspect ratio has a surface area that scales with the height squared, while the volume scales as the height cubed. For a given target xenon mass, the detector height scales with $m^{1/3}$. Since Rn emanates from the wall materials, the measured rates of radon decays should follow the ratio of surface area to volume which evolves according to $m^{-1/3}$. The red dashed line extrapolates from XENON10 and XENON100 radon levels according to the $m^{-1/3}$ model. Adapted from~\cite{rn_rate}.}
    \label{fig:radon}
\end{figure}

Ultimately, $^{222}$Rn becomes the primary contaminant since it is the hardest to mitigate and it is present in most materials as a daughter of ubiquitous $^{238}$U. Although usually present in lower concentrations and despite decaying away faster, $^{220}$Rn poses the same challenges. The beta decay spectra of their respective daughters, $^{214}$Pb and $^{212}$Pb, span the entire low-energy recoil range. The decay has characteristics of an ER event, so the S2 vs S1 ratio discrimination is vital to reducing the background in the WIMP NR region of interest. Figure~\ref{fig:radon}, shows radon rates as detectors increase in size. An activated carbon trap~\cite{LUX:2016wel}, rigorous cleaning protocols, and strict material radio-purity selections lead to a slight improvement over trends for LZ~\cite{LZ:2020fty}, which significantly improved over LUX~\cite{Bradley:2015ina}. In addition to their own radio-assay program and cleanliness~\cite{XENON:2021mrg}, XENONnT employs another distillation column dedicated to radon~\cite{Murra:2022mlr}. Online distillation of the gaseous xenon has enabled XENONnT to achieve a record background of 1.7~$\mu\mathrm{Bq/kg}$ in the initial science data, which was subsequently dropped below 1~$\mu\mathrm{Bq/kg}$ when the radon column was incorporated into the liquid xenon purification loop. Strikingly, the XENONnT ER spectrum has reduced radon so much that the lower part of the $^{136}$Xe double-beta decay spectrum is starting to dominate the ER event rate~\cite{XENON:2022ltv}.

This incredibly low radon rate is not sufficient for the proposed liquid xenon observatory accumulating a kilotonne$\times$year exposure. On one hand, since the radon emanates from detector materials surrounding the xenon target volume, the radon contamination may naturally decrease as the detector size increases, keeping all practices the same. The ratio of the detector's surface area to its volume decreases proportionally to mass contained in the volume $m^{-1/3}$, as shown in Figure~\ref{fig:radon}. However, scaling the efforts of XENONnT would still result in millions of radon ER events, of which potentially thousands would leak into the WIMP NR region of interest. A conservative goal for DARWIN of 0.1~$\mu\mathrm{Bq/kg}$ for an early 14~tonne fiducial volume results in over 100 radon ER events per year~\cite{Baudis:2013qla}. To only see 100 radon ER events in a kilotonne$\times$year exposure would require on the order of~$\sim$nBq/kg. Coating materials to prevent emanation appears to be very promising~\cite{Joerg:2022}. With fewer than one radon event per day and a good understanding of xenon fluid motion, it may be possible to remove $^{214}$Pb candidates by tracking the decay chain of individual $^{222}$Rn atoms and daughters~\cite{Qin:2023}. Thus, a multifaceted approach to reducing radon is ideal: using the purest materials with the least radon, preventing and actively removing emanated radon, and cutting radon events from analysis.

The principle material coating the inside of these detectors is polytetrafluoroethylene (PTFE). It enables the near perfect reflection of vacuum ultraviolet xenon scintillation photons, but it is porous and can easily introduce electronegative impurities, which are discussed in the next section (Section~\ref{sub_m:electronegative}), such as oxygen into the xenon via outgassing. More problematically, it can trap radioactive impurities, which can alpha decay in the PTFE and subsequently create neutrons~\cite{Cichon:2020ytl}. High standards of cleanliness significantly reduce impurities introduced by the PTFE~\cite{XENON:2021mrg,LZ:2020fty}. Since neutron NRs are a problematic background for WIMP dark matter searches and neutron capture can produce radioactive isotopes, it is worthwhile to consider investigating other reflective materials, or reducing the amount of PTFE to the minimum requirements for reflection~\cite{NEXT:2020pbx}.

\subsection{Non-Radioactive Impurities}\label{sub_m:electronegative}

Other impurities in the xenon also hurt detector performance and sensitivity. Particularly, electronegative impurities, such as diatomic oxygen, capture electrons in the drift column, detracting from S2s. Air becomes trapped in materials during detector construction and then emanates out of the materials during detector operation. As mentioned, PTFE is a major contributor. Events deeper in the detector, and therefore with longer drift times, have lower survival probabilities for electrons, which are drifting past more electronegative impurities. Typically, the purity with respect to electronegative impurities is described by the electron lifetime, which is the exponential rate constant by which electrons are lost depending on how deep in the detector the interaction occurred. For $N$ electrons initially in a cloud, if they have a drift time of $t_d$ and the electron lifetime is $\tau_e$, then the electrons that are measured (assuming perfect extraction efficiency) $N_{S2}$ are given by Equation~\ref{elifetime}.

\begin{equation} \label{elifetime}
    N_{S2} = N e^{-\frac{t_d}{\tau_e}}
\end{equation}

\begin{figure}
    \centering
    \includegraphics[scale=0.5]{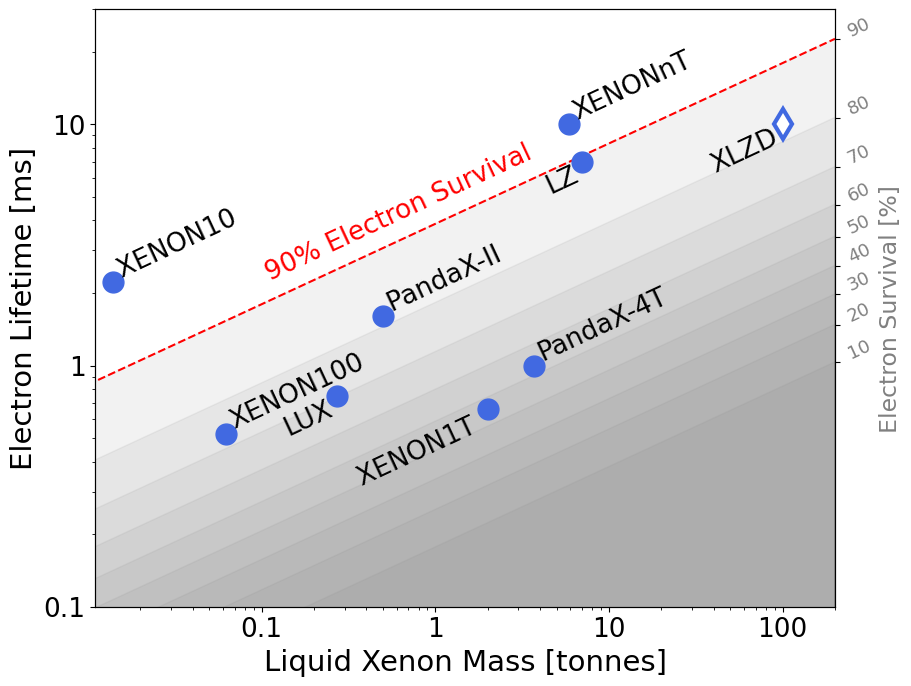}
    \caption{The electron lifetimes of current and previous detectors (blue dots) with detector mass. As mass increases, the detector gets larger and the drift times become longer. The grey shaded bands outline deciles for the survival probability of electrons at the bottom of the detector, assuming a drift velocity of 1.7~mm/$\mu$s and a 250~V/cm drift field~\cite{EXO-200:2016qyl}. The red dashed line denotes the 9th decile, above which over 90\% of electrons from the bottom of the detector survive. A blue-outlined diamond for XLZD is added, assuming the same electron lifetime as XENONnT. At lower drift fields, the slower drift velocities necessitate longer electron lifetimes for the same survival probability, and the shaded regions shift upward. Data points gathered from \cite{XENON:2010xwm,XENON:2016sjq,LUX:2020vbj,PandaX:2020oim,XENON:2021qze,PandaX:2021bab,XENON:2022ltv,LZ:2022ufs}.}
    \label{fig:elife}
\end{figure}

The strict protocols undertaken by LZ achieved an excellent electron lifetime of over 5~ms, despite only purifying through the gas phase at a rate of 3.3~tonnes/day~\cite{LZ:2022ufs}. XENONnT has developed a liquid xenon purification system~\cite{Plante:2022khm} that is able to handle 8.3~tonnes/day, operating with an electron lifetime above 10~ms~\cite{XENON:2022ltv}. Figure~\ref{fig:elife} shows the electron lifetimes recorded for previous LXe TPCs. For an example drift field of 250~V/cm in the ideal range~\cite{LUX:2020car,Szydagis:2022ikv}, the current XENONnT electron lifetime is sufficient for over 80\% of electrons at the bottom of a future 100~tonne liquid xenon observatory to reach the liquid surface.

In terms of sensitivity, scaling the current technology can ensure excellent electron lifetime and S2 detection capabilities. However, there are more subtle and dangerous consequences of non-radioactive impurities. XENON100 showed that small isolated S2 rates, particularly single-electrons, increase with more electronegative impurities~\cite{XENON:2013wdu}. Other studies into small S2s support this conclusion, although the dependence of rates on the electron lifetime is complicated~\cite{XENON:2021qze,LUX:2020vbj}. There are also non-radioactive impurities that do not capture electrons, thus not contributing to the electron lifetime calculation, which are harder to quantify. These might absorb photons, and even emit electrons as photoemitting  impurities. The stability of the S1 light yield is important for monitoring these photoemitting impurities, but their relationship to isolated S2s is not established. After a high energy event with a bright S2, single electrons continue to be measured at high rates for a maximum drift time. In XENON100, some of these electrons were directly linked to electrons emitted via the photoelectric effect from copper field-shaping rings on the inside wall of the detector~\cite{XENON:2013wdu}. The elevated rates also appeared to originate in the bulk xenon, based on timing and $(x,y)$-position. Such photoionization has largely been ignored because it can be cut with a time veto after an event that reduces exposure negligibly, depending on event rates. However, the junk data can burden the data acquisition (DAQ) and extra electrons soon after an S2 might be grouped into the S2 and negatively affect energy resolution.

Beyond a maximum drift time, there are still small S2s at elevated rates correlated with previous interactions~\cite{XENON:2021qze,LUX:2020vbj,Kopec:2021ccm}. Understanding these is vital to reducing incorrectly paired S1s and S2s that form accidental coincidence (AC) events. Discussion of such accidental backgrounds is continued in Section~\ref{ac_discuss}. For a kilotonne$\times$year exposure, controlling electronegative impurities is very important so that electrons deep in a larger detector are able to be measured. It is also critical for background suppression, but the electron lifetime parameter is proving insufficient for judging xenon purity. Further work is required to characterize and mitigate non-radioactive impurities beyond electronegative impurities, particularly photoemitting impurities, which are expected to have significant effects on backgrounds in the form of AC events and put an unnecessary burden on the DAQ.

\subsection{Background Rejection}\label{sub_t:discrimination}

As xenon detectors scale toward 100~tonnes, the characterization and suppression of backgrounds becomes more important to cover the most parameter space. The outer detectors must have a high detection efficiency for muons and their effects, namely also a high detection efficiency for neutrons. As previously mentioned, high rates of radiogenic ER events result in significant leakage into NR signal regions, otherwise obscuring potential rare processes that indicate new physics. While these listed backgrounds can be modeled and anticipated, little is known about AC backgrounds; events created from the random pairings of unrelated S1s and S2s.

Neutrons are more likely to multiply scatter in a larger detector, but it is important to characterize their background contributions, since neutrons are the only expected NR backgrounds beyond neutrino CEvNS to WIMP searches. The LZ outer detector has achieved an 89\% tagging efficiency in the Gadolinium-loaded liquid scintillator~\cite{LZ:2022ufs}. The type of rock overburden above the experiment has an impact on ambient neutron and muon rates. Going deeper underground is better to avoid muons, which are also a main contributor to free neutrons. PANDAX-4T in China's Jinping Underground Laboratory experiences a muon flux of $3.5\times10^{-10}$~cm$^{-2}$s$^{-1}$~\cite{PandaX:2021lbm}, while XENONnT in Italy's Gran Sasso National Laboratory experiences a muon flux about two orders of magnitude higher. Even so, XENONnT estimated their neutron background to be about 1 event/tonne/year with only a 53\% tagging efficiency~\cite{XENON:2023sxq}. With better neutron tagging and a larger target, the currently achieved neutron rates are acceptable, with good characterization for a kilotonne$\times$year exposure.  

All methods are worth considering to reduce radioactive ER backgrounds such as from $^{222}$Rn. Cleanliness, online removal systems and offline tagging have already been discussed. Additionally, the ability to independently consider ERs (such as for pp-solar neutrino physics and neutrinoless double-beta decay) and NRs (such as for WIMP and CEvNS searches) is important to maximize a future liquid xenon observatory's science reach~\cite{Aalbers:2022dzr}. As with argon, S1 pulse-shape discrimination is an option in xenon. However the singlet state, which dominates ERs, and the triplet state, which is higher for NRs, have time constants that are about 4~ns and 27~ns, respectively~\cite{XMASS:2018hil}. These time constants are fast compared to the typical 10~ns samples of the readout electronics, and lead to a leakage fraction of $4\times10^{-3}$~\cite{LUX:2018zdm}. Unless even better operating conditions and analysis techniques can further improve the ER/NR discrimination, achieving the optimal drift field and reducing ER rates to irreducible neutrino and xenon isotope decay backgrounds must be a major focus for a future experiment.

\subsubsection{Accidental Coincidences}\label{ac_discuss}
A challenging background to low-energy searches that is becoming more relevant for larger detectors is accidental coincidences (ACs). LXe TPCs have observed ambient rates of unpaired S1-like signals and S2-like signals, skewed to lower energies~\cite{XENON:2019izt}. Statistically, the probability of unrelated S1s and S2s being matched into events depends on the rates of these lone S1s ($R_{lS1}$) and lone S2s ($R_{lS2s}$) and the drift time ($t_d$) of the detector. Equation~\ref{acss} gives the form for the accidental coincidence rate~\cite{PandaX-II:2022waa}. Interaction quality selections ensure that most ACs are not considered signal, but the fraction of AC events that pass these data cuts, $\eta$, becomes more significant for larger exposures.

\begin{equation}
R_{AC} = \eta R_{lS1} R_{lS2} t_{d}
    \label{acss}
\end{equation}

\begin{figure}
    \centering
    \includegraphics[scale=0.5]{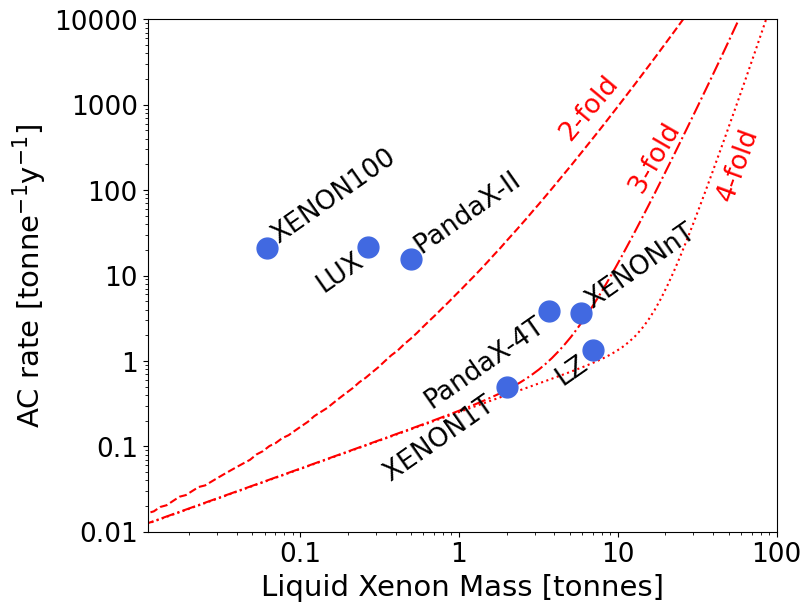}
    \caption{Estimations of the AC rates of current and previous detectors~\cite{XENON:2011cza,LUX:2016ggv,PandaX:2020oim,XENON:2018voc,PandaX:2021bab,XENON:2023sxq,LZ:2022ufs}. The red lines follow an AC model outlined by Equation~\ref{acss} that is roughly scaled to XENON1T. Three lone S1 rate scenarios were explored by setting the lowest-energy S1 threshold to include 2-fold (dashed), 3-fold (dash-dotted) and 4-fold (dotted) coincident contributing PMTs. Considerations for the model are described in detail in the text. }
    \label{fig:accidentals}
\end{figure}

There are a few identified contributions to the lone S1 rates. Some come from interactions where the ionization electrons are not measured. With higher electron lifetime, fewer interactions could lose all their ionization electrons to electronegative impurities. Additionally, an optimization is required between distancing the bottom PMTs from the increasingly high voltage of the cathode, and reducing the volume below the cathode, where ionization electrons drift down and cannot be measured. 

Depending on the single electron gain and the height of the gas amplification region, electrons could create narrow signals that may be misclassified as S1s. As discussed previously, there are high rates of single electrons for long periods after interactions~\cite{XENON:2021qze,LUX:2020vbj,Kopec:2021ccm}. In optimizing the liquid level and gate to anode spacing for the extraction electric field, the ability to distinguish single electrons from S1s should be a consideration to avoid high rates of false lone S1s with energies around the single electron gain value. Improved signal characterization algorithms between S1s and S2s, and better signal quality selection criteria, can also reduce this lone S1 contribution, complementing hardware efforts. 

High rates of the smallest S1s are mostly attributed to the random pile-up of photons or photon-like signals in the PMTs. Recently, a background of single-photons has also been observed~\cite{Qi:2023bof,XENON:2021qze,LUX:2020vbj}. It is not known how these rates might scale with detector size, but they appear connected to the overall interaction rate in the detector. Photosensor dark counts are where sensors measure a photon-like signal despite no physically incident photons. The current detectors use Hamamatsu R11410 PMTs~\cite{Antochi:2021wik,LopezParedes:2018kzu,Yang:2021hnn}, which have a typical dark rate of about 40~Hz~\cite{LopezParedes:2018kzu}. For larger detectors, the higher number of photosensors leads to a higher probability that two photons or photon-like dark counts are observed within the photon time-of-flight distribution for the detector, which would be on the order of 100~ns for meter-scale detectors.

The lone S2 background is not well-characterized. Since electrons have a much higher detection efficiency than individual photons, many low-energy events are analyzed in an S2-only channel. However, there are also S2-like signals where the origin of the measured electrons is unlikely to be a direct particle interaction. The cathode emits groups of electrons at a non-negligible rate~\cite{XENON:2019gfn}. LUX observed electron bursts after events that may be attributed to poor extraction efficiency and charge buildup at the liquid surface~\cite{LUX:2020vbj}. For S2s containing fewer than 5~electrons, there is a delayed electron background that continues for seconds after the previous events that cause them~\cite{XENON:2021qze,Kopec:2021ccm}. Better electron lifetime and low detector event rates begin to mitigate these lone S2 rates, but more characterization is needed to give a satisfactory estimate of how the AC rates scale in a larger detector and overall detector configuration.

Figure~\ref{fig:accidentals} shows the AC background rates of previous and current experiments and includes three red lines from a model based on Equation~\ref{acss}. The red lines correspond to three lone S1 rates $R_{lS1}$ depending on how many PMTs must measure photons to identify the smallest S1. They are scaled to XENON1T, which had a three-fold PMT coincidence requirement. The parameter $R_{lS1}$ includes pile-up based on an n-fold PMT coincidence requirement within 100~ns for PMTs with a 40~Hz dark count rate and an ambient photon rate depending on the event rate ($R_{\mathrm{event}}$). The event rate for XENON1T was about 5~Hz~\cite{XENON:2021qze}, corresponding to a photon rate in 248~PMTs~\cite{XENON:2018voc} of about $10^{4}$~Hz~\cite{XENON:2021qze}. The ambient photon rate was therefore chosen to scale with $10R_{\mathrm{event}}$. It also has a weak ($10^{-3}R_{\mathrm{event}}$) component that depends linearly on the event rate for real events that are missing S2s, or are single electrons misclassified as S1s. The weak dependence was chosen to give the XENON1T $R_{ls1}$ value of around 1~Hz, and the parameter $R_{lS2}$ was taken from XENON1T as 2.6~mHz~\cite{XENON:2018voc}, which was about $10^{-3} R_{\mathrm{event}}$. The lone S1 and S2 rates are assumed to depend on event rate, which is dominated by radioactivity of the materials surrounding the walls and should scale with mass as $m^{2/3}$. The total number of photosensors also scales with $m^{2/3}$. The drift time scales with $m^{1/3}$, using the drift velocity 1.7~m/ms. This leaves an $\eta$ of around 1\% to scale the model to XENON1T.

There are several differences between experiments. The typical photosensor dark rate has decreased with better technology, so the event rates per kilogram have decreased faster than $m^{-1/3}$. Also, analyses have become more complicated to select against ACs, reducing $\eta$ in Equation~\ref{acss}. Most critically, the dependence of the lone S1 and S2 rates on event rate and other parameters is not understood, and only roughly estimated here. As can be seen, the number of photosensors and their dark rates do eventually dominate the lone S1 rate.

To suppress the high AC rates, it makes sense to require an additional PMT to contribute to a measured S1 and reduce the probability that an S1 is from pile-up dark counts. Current and previous experiments have used 3-fold coincidence, except LUX, which used 2-fold and even single-photons~\cite{LUX:2019npm}. However, requiring more photons across more channels worsens the sensitivity to low-energy interactions. Particularly, there is a significant drop in sensitivity to CEvNS signals from solar $^8$B neutrinos~\cite{XENON:2020gfr} when requiring more measured photons in the S1.

The model in Figure~\ref{fig:accidentals} based on Equation~\ref{acss} assumes the event rate is dominated by the radioactivity of the materials around the detector and scales with $m^{2/3}$. The leakage into analysis $\eta$ is a constant scaled to XENON1T~\cite{XENON:2018voc}, despite expected improvements in analysis techniques. The goals for reducing $^{222}$Rn rates are expected to reduce AC rates by reducing overall detector event rates and the electron and photon backgrounds that are correlated with each event. Thus, this limited understanding of the AC rates indicates that they could be reduced to tolerable levels in a future experiment by reducing the overall event rates, which is a requirement anyways.

\subsection{Calibrations}\label{sub_t:calibration}
The detector response must be calibrated to NRs and ERs to differentiate signals for different science goals. NRs are typically calibrated with externally-applied point sources, and ERs usually use dissolved sources.

The ER response of LZ is calibrated with dissolved tritium~\cite{LZ:2022ufs}, while XENONnT and PANDAX-4T use injected $^{220}$Rn. Maintaining the same concentrations of dissolved sources when scaling up can lead to an unmanageable event rate in a larger detector. There would be event pile-up, making it difficult to distinguish which S1s go with which S2s in a given amount of time. The type of source also matters. $^{220}$Rn calibrations produce high-energy alphas that can limit the usable concentrations of the source without initially blinding the photosensors. The calibration signal, $^{212}$Pb, beta decays with a half-life of about half an hour after its parent $^{216}$Po decays. Due to xenon flow and metallic $^{212}$Pb plating onto surfaces, there will always be more measured $^{216}$Po and $^{220}$Rn alphas than the desired $^{212}$Pb beta decays. On the other hand, tritium is the more efficient calibration source, since it purely produces tritium beta decays. The trade-off is that $^{212}$Pb calibrates a larger energy range. Also, tritium can easily contaminate the system, and the first results of PANDAX-4T contain significant levels of tritium~\cite{PandaX:2021bab}. Calibrating the ER response of a future detector warrants consideration for other beta radiation sources. With a radon distillation column, $^{222}$Rn may be itself used as a calibration source to characterize the detector against its natural background levels. Its decay chain takes more time, allowing the photosensors to be off when bright alpha rates are high, and then take data while $^{214}$Pb betas dominate at later times.

The NR response of LZ is calibrated with both an external deuterium-deuterium (DD) neutron generator aimed at the cryostat and an external $^{241}$AmLi point source to additionally measure the outer neutron detector efficiency~\cite{LZ:2022ufs}. PANDAX-4T is also calibrated with an external DD neutron generator for NR~\cite{PandaX:2021bab}. Meanwhile, XENONnT uses a $^{241}$AmBe neutron source near the cryostat, which also calibrates the outer neutron veto detector~\cite{XENON:2022ltv}. A disadvantage of external neutron sources is that events will be localized outside the fiducial volume near the source along the edge of the detector where the electric field is less uniform. Neutrons are more likely to multiple-scatter in a large detector, so it may be necessary to split multiple-scatters into individual single-site scatters via analysis techniques to increase statistics. As an example, two S2s clearly clustered in two distinct locations can be analyzed separately. For a target of roughly a hundred tonnes of xenon, novel solutions for point sources would optimize calibration statistics. For example, if neutrons were transmitted through a tube that could be inserted into the center of the xenon target, they would calibrate the fiducial volume directly. The practicality of such a neutron insertion device would need to be rigorously investigated. 

All three current experiments calibrate position reconstruction algorithms and variations across the detector volume with dissolved $^{83m}$Kr~\cite{Kastens:2009pa}. XENONnT additionally calibrates low-energy position dependencies by injecting $^{37}$Ar, since it can be removed by adjusting the configuration of the krypton distillation column~\cite{XENON:2022ivg}. Both $^{83m}$Kr and $^{37}$Ar may still be well-suited to a future detector. However, the usable rates of any source in a detector will decrease as the volume increases to avoid pile-up. This means that accumulating statistics for precision will require these calibrations to last longer at lower dissolved concentrations, which can be accommodated. Otherwise better analysis methods of disentangling overlapping S1s and S2s must be developed. As discussed, current external NR calibrations may be impractical for such a large volume, and improvements in ER calibrations is also necessary.

\subsection{Photosensors}\label{sub_t:photosensors}

The current generation of PMTs have a high failure rate for an experiment that needs to run for over a decade. Recent protocols have improved tests of PMT quality to choose ones less likely to fail. For example, elevated levels of argon in a PMT predict a higher failure rate due to vacuum leaks~\cite{Antochi:2021wik}. In XENONnT, 17 of 494 PMTs were not working before the end of the first year~\cite{XENON:2022ltv}. This 3\% failure rate includes commissioning, which is the most stressful time for the PMTs. The passive failure rate of the good surviving PMTs is expected to be lower. However, depending on the true typical rates of vacuum leaks, failure rates are still higher than acceptable for the lifetime of a future xenon observatory. 

Advantages of PMTs include high quantum efficiency, low dark count noise rates and the 20\% probability to measure a double-photoelectron signal for xenon scintillation light~\cite{Akerib:2021pfd,XENON:2020gfr}. Their timing response is also short compared to a photon's time-of-flight. However, recent developments in silicon photomultipliers (SiPMs) indicate that they may be a more reliable choice~\cite{Baudis:2018pdv}. While PMTs do not work well in electric fields, SiPMs would be able to completely tile the inside of a detector for 4$\pi$ coverage~\cite{DARWIN:2022ewl}. Hybrid detectors, such as ABALONE~\cite{DAndrea:2021fro}, leverage some advantages of both. There may also be promise for other Single-Photon Avalanche Diodes (SPADs). It is clear that photosensors are a high community priority to maximize quantum efficiency and durability while minimizing intrinsic radioactivity and dark count backgrounds. 

\subsection{Electrodes}\label{sub_t:electrodes}

\begin{figure}
    \centering
    \includegraphics[scale=0.42]{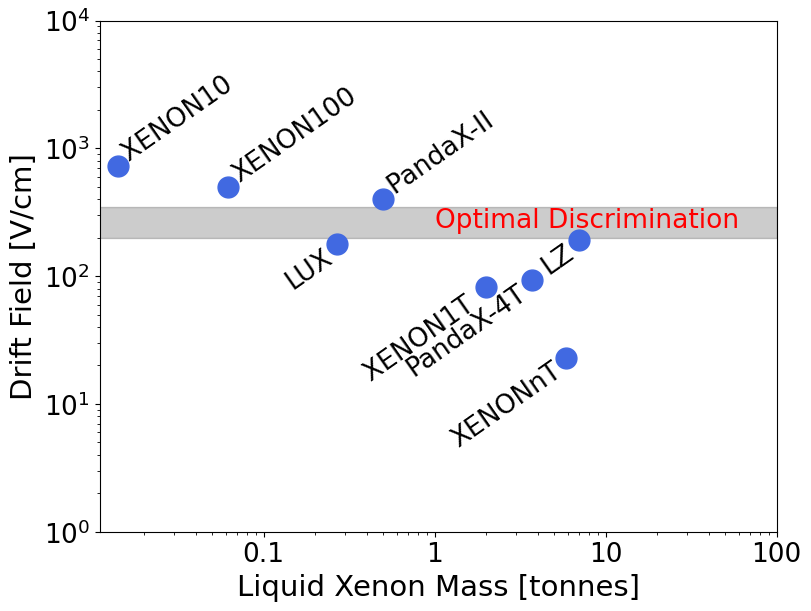}
    \includegraphics[scale=0.42]{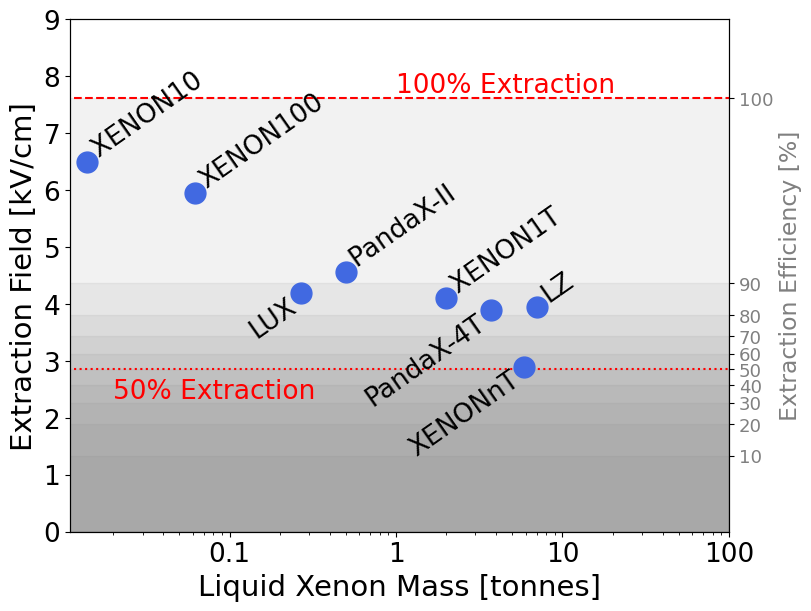}
    \caption{In scaling up in mass, LXe TPCs have seen a reduction in relative performance of electrodes. Data points gathered from \cite{XENON:2010xwm,XENON:2016sjq,LUX:2020vbj,PandaX:2020oim,XENON:2021qze,PandaX:2021bab,XENON:2022ltv,LZ:2022ufs}. \textbf{Left:} The drift fields of current and previous experiments (blue dots) with a typical optimal field for NR/ER discrimination as determined in~\cite{LUX:2020car,Szydagis:2022ikv} (grey band). \textbf{Right:} The extraction fields of current and previous experiments (blue dots) with grey shaded bands of the deciles of electron extraction from~\cite{Xu:2019dqb}. The 100\% and 50\% decile lower limits are marked with red horizontal dashed and dotted lines respectively.}
    \label{fig:fields}
\end{figure}

In order to measure S2s, the electric fields are critical. Figure~\ref{fig:fields} shows that the typical operating fields of experiments are decreasing. This is particularly problematic for the extraction field, since the electron extraction efficiency has an abrupt roll-off~\cite{Xu:2019dqb}. The drift field also has significant importance for discriminating between ER backgrounds and NR signal. With larger detectors and larger electrode spacing, higher absolute voltages are required to create the same field strengths. For a 100~tonne detector and equal aspect ratio, the drift region would be about 3~m, which would require a potential difference of roughly 75~kV for a drift field of 250~V/cm. Achieving an extraction field of 7.5~kV/cm in the liquid would require optimization of the gate-anode spacing and the liquid level above the gate. Lower potential differences are required for smaller spacing, and when the liquid xenon level is a higher fraction of the space between the gate and anode. However, that negatively affects the proportional scintillation yield with a smaller gas gap.

The target electric fields are well motivated for a LXe TPC. However, there is significant room for improvement to curtail current trends toward lower fields. For electrodes 3~m in diameter, the transparency and mechanical strength have to be optimized so that S1 light reaches photosensors, and ionization electrons proportionally scintillate. The woven mesh electrodes of LZ appear to be performing the best among the current generation of detectors~\cite{LZ:2022ufs}. Alarmingly, XENONnT sporadically observes high rates of single electrons at highly localized positions that lead them to ramp down the anode, despite operating at low field configurations~\cite{XENON:2022ltv}. Electrodes with large diameters would also be under electrostatic strain to distort in the middle, creating field variations across the detector. The anode in the gas sags down due to gravity and electrostatic strain. The gate and cathode in the liquid experience less gravitational strain due to buoyancy, and actually bow upward with electrostatic strain. 

There is another option to consider: proportional scintillation in the liquid phase. A single-phase TPC has more thermodynamic stability and liquid xenon is a better insulator than gaseous xenon. One tested design consists of a radial detector with photosensors around the barrel and a single anode wire down the middle~\cite{Qi:2023bof}. Fiducialization is based on drift time to the center, and vertical position as determined by the photosensors. So far, a reasonable single electron gain has not yet been achieved to be distinguishable from single photon signals. Also, the detector observed high rates of such photon signals. It is unclear whether such a detector could easily scale over four orders of magnitude in mass for a liquid xenon observatory, but the technical challenge of enormous planar electrodes warrants the exploration of creative solutions.

\subsection{Long-Term Considerations}\label{sub_m:durability}

A difficult challenge to a future liquid xenon observatory is to run stably for over a decade. The failure rates of pumps, electronics, and stressed components must be evaluated in shorter time windows and usefully extrapolated to the lifetime of the experiment. Many components in the purification loops must be accessible and have redundancies, enabling regular maintenance. The ability to perform maintenance on the xenon TPC components with minimized down-time will need to be considered. All surfaces in contact with xenon must be thoroughly scrubbed of impurities. Also, the cryostat and pipes must contain the system without leaks. Outer detectors that tag neutrons and muon-induced events are typically based in liquid: concentric muon and neutron vetoes in water in XENONnT~\cite{XENON:2023sxq} and a liquid scintillator neutron and water muon outer detector in LZ~\cite{LZ:2022ufs}. The stability of the purification, containment, and instrumentation of these detectors requires careful consideration, particularly if the xenon cryostat also needs to be accessible for maintenance.

With an increase in detector size and number of photosensors, the DAQ will need to handle large quantities of data with a timing resolution on the order of nanoseconds for many years. Other experiments, particularly at the Large Hadron Collider, have systems in place to deal with large amounts of data, and a future LXe observatory data system is still within the scope of current technology. The failure rates of power sources and readout electronics would need to be estimated, but they are expected to already be sub-percent annually, since no failures in the current detectors have been reported. 

The energy requirements to power such an experiment are sizable. The detector monitoring systems, DAQ, pumps, high voltage sources of the electrodes and photosensors all consume large amounts of electricity. As the effects of global warming become widespread, conscientious energy practices are an ethical imperative. 

Liquid xenon necessitates temperatures of around $-100^{\circ}$C, which are currently maintained by pulse tube refrigerators (PTRs), as in XENONnT~\cite{XENON:2021mrg}, or liquid nitrogen, as in LZ~\cite{LZ:2015kxe}. As seen in the XMASS detector, PTRs can get clogged and become less effective~\cite{XMASS:2018koa}. Therefore, liquid nitrogen cooling is optimal, although it requires about 550L/day as estimated for a 50~tonne experiment~\cite{Giboni:2019fqo}. Ensuring that the 100~tonne monolithic detector volume remains in thermodynamic stability with a gas layer presents a particular challenge. The liquid and gas temperatures as well as the gas pressure directly affect the liquid xenon density, complicating efforts for precise control over the liquid surface with millimeter precision. Slight changes can have large effects that take long times to re-equilibrate throughout the system.

Once an optimal configuration is reached, the detector requires experts to maintain a stable operation. Science driven by academia has high turnover rates in personnel. The tenure of American PhD students averages 6~years~\cite{Mulvey:2021ghj}, but the time is shorter for students in other countries who typically complete masters and doctoral degrees separately. Postdoc positions are usually 2-3~years. Tenured and tenure-track faculty have other obligations that hinder their direct involvement in building, maintaining, and characterizing an experiment. For an experiment to work continuously for over a decade, permanent research personnel must be hired to ensure smooth operation without jarring transitions and loss of critical expertise. There must also be exciting work to support several doctoral theses along the way to acquiring the full kilotonne$\times$year exposure. Periodic analysis updates ensure the project remains relevant and promotes continuity.

\section{Summary and Outlook}\label{summary}

This work has discussed several aspects of current and previous experiments that cannot directly scale to a future liquid xenon observatory without improvement. In order to reach the best dark matter sensitivity, the detector materials must be carefully considered, and technological aspects of the detector and supporting systems must be improved. Acquiring the xenon requires a well-considered timeline over several years in addition to the experiment stably running for over a decade.

For discovery potential with a liquid observatory, backgrounds must be mitigated. The monoenergetic peaks of $^{124}$Xe and the double-beta decay spectrum of $^{136}$Xe are able to be constrained and subtracted, but they still have larger uncertainties than required for a kilotonne$\times$year exposure. As with previous generations, $^{222}$Rn will continue to be the major threat that requires a multifaceted approach. It is necessary to first reduce initial radon levels in the materials with strict screening and cleaning procedures~\cite{LZ:2020fty,XENON:2021mrg} and limit emanation with coatings~\cite{Joerg:2022}. Then, emanated radon must be removed via online distillation~\cite{Murra:2022mlr} and trapping~\cite{LUX:2016wel}. Finally, radon background events can be cut in analysis~\cite{Qin:2023}. Optimal detector operation must be achieved regarding photosensor dark count rates and electric fields, maximizing ER/NR discrimination. New analysis techniques must be pursued to select signal candidates and reject backgrounds, particularly AC. Machine learning presents an attractive opportunity. For example, a Bayesian network was applied to XENONnT data to discriminate S1s and S2s as a proof of concept that would reduce misclassification in lone S1 and S2 rates~\cite{XENON:2023dar}. Such isolated S1 and S2 rates are alarmingly high in current experiments. Combined with longer drift times, they exacerbate AC backgrounds. Significant work in current systems is required to better anticipate AC backgrounds and how they scale with detector size and configuration.

As the community moves to the ultimate liquid xenon observatory, there is a need for greater transparency regarding all challenges of the current generation. Nuclear recoil calibrations as they are currently conducted from outside the detector are not sustainable for statistics and full detector characterization. Rigorous testing of photosensors is necessary to mitigate failure risks and determine durability. Open discussions are vital to determining the best hardware for long-term stable operations and lowest impurity contamination. PTFE requires strict cleanliness protocols, and other options may be available to reduce its effect on radioactive and non-radioactive impurities alike. Most crucially, electrode geometry, material and treatment must be optimized and behaviors characterized to halt the trend of worsening electric fields with larger experiments.

The wealth of science opportunities of a future liquid xenon observatory~\cite{Aalbers:2022dzr} invites the participation of the whole LXe community. In the coming decades, such a project promises rich results spanning many interesting fields. The technological challenges, along with the research and development still required to overcome them, have wider applications beyond WIMP dark matter searches. For example, there is the exciting possibility to measure CEvNS from the next galactic supernova~\cite{Lang:2016zhv}. Additionally, the guaranteed measurement of CEvNS from solar neutrinos has momentous astrophysical implications, and neutrinoless double-beta decay searches probe the very nature of neutrinos. While scaling up in mass comes with growing pains, LXe TPCs are well-positioned to overcome these challenges and continue their successful lineage.

\section*{A\lowercase{cknowledgments}}

The author thanks the XENON, LUX-ZEPLIN, and DARWIN Collaborations in the XLZD consortium for vital discussions and preliminary guidance. Particularly, I acknowledge Dr. Tina Pollmann for expertise in the contributing factors to Accidental Coincidence. This work was partially supported by the National Science Foundation Mathematical and Physical Sciences Ascending (MPS-Ascend) Postdoctoral Research Fellowship 2137911.
    
\bibliographystyle{JHEP}
\bibliography{bibliography} 

\end{document}